\documentclass[prl,twocolumn,preprintnumbers,amsmath,amssymb]{revtex4}
\usepackage{graphicx}
\usepackage{bm}
\usepackage{hyperref}
\usepackage{epsfig}
\usepackage{amsfonts}
\usepackage{amsthm}
\usepackage[T1]{fontenc}

\begin{document}

\title{Time-Loop Formalism for Irreversible Quantum Problems: Steady State Transport in Junctions with Asymmetric Dynamics}

\author{P. Kakashvili}
\author{C. J. Bolech}
\affiliation{Physics \& Astronomy Department, Rice University, Houston, TX
77005, USA}


\begin{abstract}
Non-unitary quantum mechanics has been used in the past to study
irreversibility, dissipation and decay in a variety of physical
systems. In this letter, we propose a general scheme to deal with
systems governed by non-Hermitian Hamiltonians.  We argue that the
Schwinger-Keldysh formalism gives a natural description for those problems. To
elucidate the method, we study a simple model inspired by mesoscopic
physics  --an asymmetric junction. The system is governed by a
non-Hermitian Hamiltonian which captures essential aspects of
irreversibility.
\end{abstract}

\pacs{72.10.Bg, 03.65.Yz, 05.70.Ln, 73.23.-b}

\maketitle

Non-Hermitian formulations of quantum mechanical problems have
attracted substantial interest in almost all areas of physics. In most
cases, non-unitary approaches are utilized to describe irreversible
processes, --such as decay and dissipation--, in open quantum
systems. This type of problems have been addressed since the early
days of quantum mechanics, when the complex eigenvalue method was
pioneered to describe the $\alpha$-decay of
nuclei~\cite{gamow_zur_1928,gurney_wave_1928}. Since then, non-unitary
approaches have been applied to theories of $K$ and $B$ meson
decay~\cite{lee_remarks_1957,winstein_search_1993}, scattering and
absorption of particles by nuclei~\cite{feshbach_scattering_1947},
nuclear reactions~\cite{feshbach_model_1954,feshbach_unified_1958},
multi-photon ionization of
atoms~\cite{baker_non-hermitian_1984,faisal_theory_1987}, optical
resonators~\cite{siegman_orthogonality_1979} and free-electron
lasers~\cite{dattoli_lethargy_1988}. Growth models have been
investigated by mapping a master equation into a Shr\"odinger equation
with a non-Hermitian Hamiltonian~\cite{gwa_bethe_1992}.  In the last
decade, non-Hermitian theories have also been applied to condensed
matter systems with localization-declocalization transitions.
Pinning/depinning of flux lines from columnar defects has been
studied~\cite{hatano_localization_1996_combined} in
type-II superconductors, and the effects of a single columnar defect on
fluctuating flux lines has been considered by mapping to a non-Hermitian
Luttinger Liquid model and using the density matrix renormalization
group (DMRG)~\cite{affleck_non-hermitian_2004}. Generalized DMRG has
been applied also to a one-dimensional reaction-diffusion model with
non-unitary time evolution~\cite{carlon_density_1999} and a
non-Hermitian spin-1/2 Heisenberg chain~\cite{kaulke_dmrg_1998}. The
closing of the Mott gap has been investigated in the non-Hermitian
Hubbard model~\cite{fukui_breakdown_1998} and evaporatively cooled
Bose-Einstein condensates (BECs) have been studied by a non-unitary
quantum dynamics approach~\cite{drummond_quantum_1999}. While all of
the above studies used a variety of case-specific methods and were 
mainly concerned with either time-independent
or transient behaviors, in this letter we present a general framework,
that we apply to a steady-state formulation of an irreversible system.
\begin{figure}[!hpb]
\begin{centering}
\includegraphics[width=3.4in,keepaspectratio]{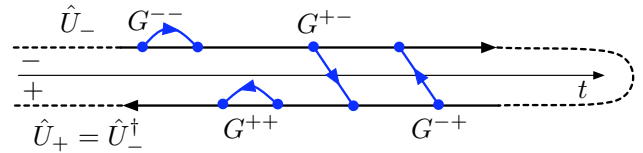} \par
\end{centering}
\caption{Sketch of the Keldysh contour. The arrows show
the direction of the time evolution. Here $G^{--}$ is the time-ordered
Green's function, $G^{++}$ is the anti-time-ordered Green's function,
$G^{+-}$ is the ``greater'' Green's function and $G^{-+}$ is the
``lesser'' Green's function.}
\label{KeldyshContour} 
\end{figure}

To define our approach we start with the
Schwinger-Keldysh (SK) formalism~\cite{schwinger_brownian_1961,keldysh_diagram_1965},
which turns out to be very natural in the non-Hermitian case. In
the SK formulation, the evolution of a system is described on a
time loop (Keldysh contour, $K$) with ``forward'' (``$-$'') and
``backward'' (``$+$'') directions (see
\protect{Fig.~\ref{KeldyshContour}}), thus defining two distinct branches for
time evolution. Special care should be taken to define proper
evolution operators for each direction. One could argue that the time
evolution operator is the same along the full path, as in the Hermitian case,
$\hat{U}_{K}(t,t_{0})=\hat{U}_{+}\hat{U}_{-}=T_{K} e^{-i\int_{t_{0}}^{t} dt' \hat{H}(t')}$ (where $T_{K}$
is the time-ordering operator along the Keldysh contour); which preserves the normalization
along the contour, $\hat{U}_{K}(t_{0}^{+},t_{0}^{-})=\openone$.  This would mean that the
time evolution would be governed by $\hat{H}$ on both branches. Even
though this choice ensures that the ``backward'' evolution is the
algebraic inverse of the ``forward'' one and it may seem a good
choice, we should emphasize that it does not describe the physical
process faithfully. Physically, in the ``backward'' branch, the system
should \emph{rewind} its ``forward'' evolution and therefore
$\hat{H}^{\dagger}$ should be the operator which dictates the
dynamics. Then the ``backward'' time evolution operator
is given by $\hat{U}_{+}(t_{0},t)=\hat{U}_{-}^{\dagger}(t,t_{0})=\tilde{T} e^{i
\int_{t_{0}}^{t} dt'
\hat{H}^{\dagger}(t')}\ne\hat{U}_{-}^{-1}(t,t_{0})$, which is
automatically anti-time-ordered (with $\tilde{T}$ representing the
anti-time-ordering operator). Of course, one recovers the familiar
result in the Hermitian case. We see that this approach would be hard
to implement in a single-time formalism and a double-time approach is
essential and rather natural for non-Hermitian problems. Thus, the
``$+$'' branch is governed by the Hermitian conjugate theory and the
time-ordered Green's function ($G^{--}$) is mapped into the
anti-time-ordered one ($G^{++}$) by Hermitian conjugation; which is
all natural in a time-loop formalism. This makes a generalized SK method an
appealing choice
\footnote{We would like to point out another formalism with a doubled
time basis, that has been applied to transients in irreversible quantum
dynamics~\cite{bohm_microphysical_1995}, in which the doubling of the
states is related to the time-reversal invariance properties and the
state-space doubling described by Wigner~\cite{wigner_group_1964_short}.}.

\begin{figure}[!ht]
\begin{centering}
\includegraphics[width=3in,keepaspectratio]{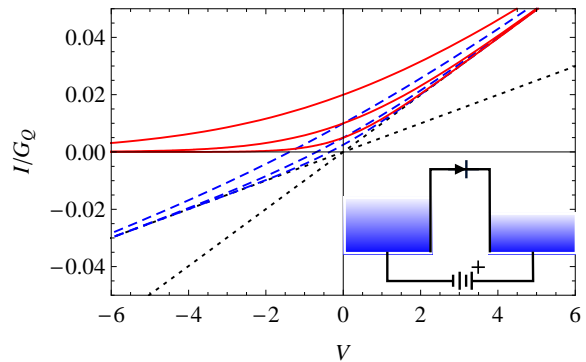} \par
\end{centering}
\caption{Current, $I$, as a function of voltage, $V$, for different temperatures (in arbitrary units). The current is shown for $t_{R}=t_{L}=0.05$ and $t_{R}=t_{L}=0.05/\sqrt{2}$ (SJ case, \emph{dotted} lines), for $t_{R}=0.05$ and $t_{L}=0.05/\sqrt{2}$ (\emph{dashed} lines) and for $t_{R}=0.05$ and $t_{L}=0$ (\emph{solid} lines). The temperature increases from bottom to top and $G_{Q}$ indicates
the quantum of conductance. \emph{Inset}: Sketch of the tunneling setup.}
\label{current} 
\end{figure}

To illustrate the formalism, we shall study a simple model of a
single-mode asymmetric tunneling junction (see inset of
Fig.~\ref{current})~\footnote{We just learned that a model of this
type was recently discussed by Datta [to appear in the Proceedings of
the Third ASI International Workshop on Nano Science \& Technology,
Ed.~Z.~K.~Tang, (Taylor \& Francis, 2007)].}, which captures essential
aspects of irreversibility. We shall study the steady state transport
of the system.  The Hamiltonian of the model is given by
$\hat{H}=\hat{H}_{0}+\hat{H}_{t}$, where
\begin{equation}
\hat{H}_{0}=\sum_{r=R,L}\hat{H}_{0,r}=-iv\sum_{r=R,L}\int dx\,\hat{\psi}_{r}^{\dagger}(x)\partial_{x}\hat{\psi}_{r}(x)
\label{eq:leads}
\end{equation}
\noindent describes noninteracting left ($L$) and right ($R$) leads in
a one-dimensional formulation~\cite{affleck_conformal_1995}.
$\psi_{r}^{\dagger}(x)$ [$\psi_{r}(x)$] represent electron creation
[annihilation] operators in the corresponding lead and $v$ is the Fermi
velocity which is taken to be the same for both leads.

The irreversible nature of the problem is given by
\begin{equation}
\hat{H}_{t}=2v \sum_{r\ne r'}t_{r}\,\hat{\psi}_{r}^{\dagger}(x \! = \! 0)\hat{\psi}_{r'}(x \! = \! 0),
\label{eq:tunneling}
\end{equation}
\noindent which describes the tunneling with rates $t_{L}\ne t_{R}^{*}$ in
the asymmetric junction (AJ) case. (In
this letter, $k_{B}=\hbar=e=1$.)

Let us derive a local theory which captures the essential physics of the
junction. After integrating out the leads exactly in a path-integral
formulation and dropping the coordinate dependence of the electron creation
and annihilation operators, the local action reads
\begin{equation}
{\cal A}=\int_{-\infty}^{\infty}\bar{\Psi}(t)\mathbf{G}^{-1}(t-t')\Psi(t')\, dtdt',
\label{eq:action}
\end{equation}
\noindent where $\Psi=\left(\begin{array}{cccc}
\psi_{L}^{-} & \psi_{R}^{-} & \psi_{L}^{+} & \psi_{R}^{+}\end{array}\right)^{T}$. We have doubled the basis of the problem, which is customary
in the SK formulation, and assigned extra indices ``$-$'' and
``$+$'' to the fields according to the particular branch of the Keldysh
contour they belong to. In Eq.~(\ref{eq:action}), we explicitly
imposed the steady-state constraint by specifying the time dependence of
the inverse Green's function $\mathbf{G}^{-1}$; In the steady state,
the Green's function depends only on the time difference, $t-t'$, and
not on the ``center-of-mass'' time
$(t+t')/2$~\cite{lake_nonequilibrium_1992,bolech_point-contact_2004_combined}. Thus, the Fourier
transform (with respect to the time difference) is given by
\begin{equation}
\frac{\mathbf{G}^{-1}(\omega)}{-2vi}= \left(\begin{array}{cccc}
-s_{L} & -it_{L} & s_{L}-1 & 0\\
-it_{R} & -s_{R} & 0 & s_{R}-1\\
s_{L}+1 & 0 & -s_{L} & it_{R}^{*}\\
0 & s_{R}+1 & it_{L}^{*} & -s_{R}  \end{array}\right),
\label{eq:invGreen}
\end{equation}
\noindent where $s_r=\tanh\frac{\omega-\mu_r}{2T_r}$. Here, we assumed
that the leads are in equilibrium at their respective
chemical potentials ($\mu_{L}$ and $\mu_{R}$) and temperatures
($T_{L}$ and $T_{R}$). 

Using the Green's functions, we can calculate
the transport properties of the asymmetric junction.

\emph{Current} - We first describe a current flowing across the junction.
We expect that there should be an asymmetry between positive and
negative voltage bias. We define the current operator in the usual way
by $\hat{I}=\frac{1}{2}\partial_{t}\Delta\hat{N}$, where
$\Delta\hat{N}=\hat{N}_{R}-\hat{N}_{L}$, and $\hat{N}_{L}$ and
$\hat{N}_{R}$ are the particle number operators for the left and right
leads, respectively. While the current is defined unambiguously for a
symmetric junction (SJ), this is not the case for an asymmetric one.
The above formula is incomplete if we do not specify on which branch
$\Delta\hat{N}$ resides. On the ``$-$'' branch, the current operator is
given by
$\hat{I}^{-}=\frac{1}{2}\partial_{t}\Delta\hat{N}^{-}=\frac{i}{2}[\hat{H},\Delta\hat{N}^{-}]$;
since $\hat{H}$ governs the time evolution on this branch.  This
prompts us to use the time-ordered (``$--$'' component) Green's
functions for calculations. The current is thus given by
\begin{eqnarray}
I^{-} = 2v \int_{-\infty}^{\infty}\frac{d\omega}{2\pi}[t_{L}G_{RL}^{--}(\omega)-t_{R}G_{LR}^{--}(\omega)].
\label{eq:current-}
\end{eqnarray}

\begin{figure}[!hpb]
\begin{centering}
\includegraphics[width=3in,keepaspectratio]{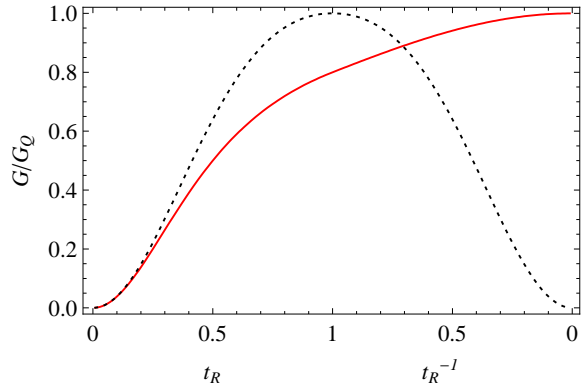} \par
\end{centering}
\caption{Large-voltage conductance as a function of tunneling
strength ($t_{R}$) for a SJ, $t=t_{R}$ (\emph{dotted} line) and in the
extreme AJ case, $t_{L}=0$ (\emph{solid} line).}
\label{conductance} 
\end{figure}

For consistency, we should get the same result if we do calculations
on the ``$+$'' branch. In this case, the current operator reads
$\hat{I}{}^{+}=\frac{1}{2}\partial_{t}\Delta\hat{N}^{+}=\frac{i}{2}[\hat{H}^{\dagger},\Delta\hat{N}^{+}]$;
in analogy with the ``$-$'' branch. Here $\hat{H}^{\dagger}$ defines
the time evolution and we should use the anti-time-ordered (``$++$''
component) Green's functions. In this case the current has the same form
as in Eq.~(\ref{eq:current-}), but with time-ordered Green's functions
replaced by anti-time-ordered ones and $t_{L/R} \rightarrow t_{R/L}^{*}$.
%
%
Direct calculations indeed show that $I^{-}=I^{+}=I$.
In the SJ
limit, calculations can be done analytically and the current is given
by $I=GV$, where $G=\frac{2|t|^{2}}{\pi(1+|t|^{2})^{2}}$
\cite{blanter_shot_2000} and $V=\mu_{L}-\mu_{R}$. The well known symmetry
of the conductance in the SJ case, $|t| \rightarrow 1/|t|$, is generalized
to $t_{R/L} \rightarrow 1/t_{L/R}^{*}$ for the AJ one. The results for
the current in the weak tunneling limit are shown on Fig.~\ref{current},
where we see the anticipated asymmetric
behavior for unequal tunneling strengths. For large voltages the
behavior is similar to the SJ case with the corresponding tunneling
rate ($t=t_R,t_L$ for positive or negative bias, respectively). In the
extreme limit when $t_{L}=0$, we see a clear diode-like behavior. The
results also show temperature dependence (that disappears in the SJ
limit), in particular, the current is nonzero for zero voltage due to
finite temperature effects ($I_{0} \propto T$). For intermediate and large tunneling
strengths, the current is significantly different from the SJ value
even for large voltages. This can be seen on Fig.~\ref{conductance},
which shows the conductance as a function of tunneling rate.

The behavior of the conductance can be understood in terms of a
perturbative expansion. For weak tunneling ($t,t_{R} \ll 1$), the
expansions for the symmetric and asymmetric cases are identical
(\emph{i.e.}, the leading orders coincide). While for intermediate
tunneling ($t,t_{R} \sim 1$), many processes which are allowed for
the SJ are not allowed for the extreme asymmetric one, hence the
reduction of the conductance. For large tunneling strengths ($t,t_{R}
\gg 1$), the conductance decreases in the SJ case due to a
resonance developing at the junction. In the AJ case, this resonance
is suppressed and, therefore, the conductance increases, and saturates
at the quantum of conductance in the limiting case
($t_{R}\rightarrow\infty$).

\emph{Noise} - Now we turn to the noise, which is an $\omega$-dependent
response~\cite{blanter_shot_2000,bolech_observing_2007}. The current-current correlation function is usually defined
in the following way
\begin{equation}
S(t,t')=\left\langle :\hat{I}(t)::\hat{I}(t'):+:\hat{I}(t')::\hat{I}(t):\right\rangle ,
\label{eq:NoiseDef1}
\end{equation}
\noindent where $:\,:$ denotes normal ordering. Bearing in mind the
SK formalism, we redefine $S$ as
\begin{equation}
S(t,t')=\sum_{\eta\ne\eta'}\left\langle
T_{K}:\hat{I}^{\eta}(t)::\hat{I}^{\eta'}(t'):\right\rangle ,
\label{eq:NoiseDef2}
\end{equation}
\noindent which is equivalent to the definition in Eq.~(\ref{eq:NoiseDef1}).

\begin{figure}[!hpb]
\begin{centering}
\includegraphics[width=3in,keepaspectratio]{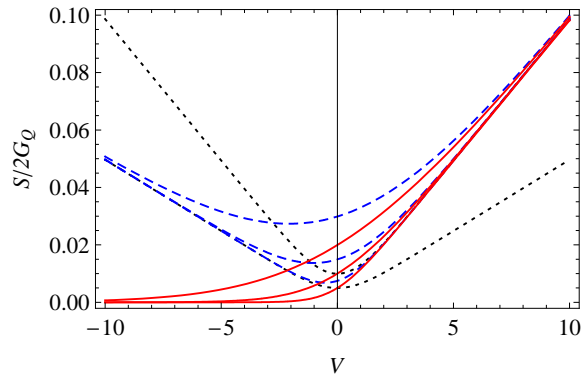} \par
\end{centering}
\caption{Current noise power, $S(\omega\!=\!0)$, as a
function of voltage, $V$, for different temperatures (in arbitrary
units). The line styles and parameters are the same as in
Fig.~\ref{current}.}
\label{noise} 
\end{figure}

From the definitions of the current operators on the corresponding
branches, using Wick's theorem \cite{wick_evaluation_1950} and Fourier transforming,
we get the current noise power
\begin{eqnarray*}
\frac{S(\omega)}{4v^2} & =  & t_{L}t_{L}^{*}\{[G_{LL}^{+-}\circ G_{RR}^{-+}](\omega)+[G_{RR}^{-+}\circ G_{LL}^{+-}](\omega)\} \nonumber \\
&+& t_{R}t_{R}^{*}\{[G_{RR}^{+-}\circ G_{LL}^{-+}](\omega)+[G_{LL}^{-+}\circ G_{RR}^{+-}](\omega)\} \nonumber \\
& - & t_{R}t_{L}^{*}\{[G_{LR}^{+-}\circ G_{LR}^{-+}](\omega)+[G_{LR}^{-+}\circ G_{LR}^{+-}](\omega)\} \nonumber \\
& - & t_{L}t_{R}^{*}\{[G_{RL}^{+-}\circ G_{RL}^{-+}](\omega)+[G_{RL}^{-+}\circ G_{RL}^{+-}](\omega)\},
\label{eq:Noise}
\end{eqnarray*}
\noindent where the correlation product is defined as
\begin{equation}
[G_{1}\circ G_{2}](\omega)=\int_{-\infty}^{\infty}\frac{d\omega'}{2\pi}G_{1}(\omega'-\omega)G_{2}(\omega').
\label{eq:CorrDef}
\end{equation}
\noindent In the SJ limit, we get the well established result \cite{blanter_shot_2000}:
\begin{equation}
S(\omega\!=\!0)=8\pi G^{2}T+2G(1-2\pi G)V\coth \frac{V}{2T}.
\label{eq:NoiseHermitian}
\end{equation}
The dependence of $S(\omega\!=\!0)$ on bias voltage and temperature,
for weak tunneling, is shown on Fig.~\ref{noise}.
The results show
similar asymmetric behavior as for the current. For large voltages,
the behavior of the noise power is similar to that of the SJ with the
corresponding tunneling rate (in exact analogy with the behavior of
the current). In the limiting case with $t_{L}=0$, we see the
suppression of the noise for negative voltage bias, which is
consistent with the diode-like behavior and is due to the reduction of
the current in this regime (see Fig.~\ref{current}). In the shot noise
limit ($V \gg T$), the Fano factor $F\equiv S(\omega\!=\!0)/(2I)=1$
for week tunneling as in the SJ case.

\emph{Heat current} - Now we shall turn to the thermal conductivity,
calculate the heat current, $I_{Q}$, and check the Wiedemann-Franz
law. To define the heat current, we use the first law of thermodynamics
to write
\begin{equation}
\text{\textit{\dj}} \Delta Q=d\Delta E+\mu_{L}dN_{L}-\mu_{R}dN_{R},
\label{eq:Term1Law}
\end{equation}
\noindent where $\Delta Q=Q_{R}-Q_{L}$ and $\Delta E=\left\langle \hat{H}_{0,R}-\hat{H}_{0,L}\right\rangle $.
By using the usual definitions for the currents, choosing a particular branch
(``$-$'') on the Keldysh contour and defining
the average chemical potential $\bar{\mu}=(\mu_{L}+\mu_{R})/2$, we get $I_{Q}^{-}=I_{E}^{-}-\bar{\mu}I^{-}$.
%
%
Using the equations of motion for the electron creation and
annihilation operators, and performing some algebra, the heat current
reads
\begin{equation}
I_{Q}^{-}=2v \int\frac{d\omega}{2\pi}(\omega-\bar{\mu})[t_{L}G_{RL}^{--}(\omega)-t_{R}G_{LR}^{--}(\omega)].
\label{eq:HeatCurrent2}
\end{equation}

Our calculations on the ``backward'' branch yield $I_{Q}^{+}=I_{Q}^{-}=I_{Q}$,
in exact analogy with the result for the electric current. In the SJ
case, we obtain the well known result \cite{sivan_multichannel_1986} $I_{Q}=K\Delta T$,
%
%
with $K=\frac{2 \pi \bar{T} |t|^{2}}{3(1+|t|^{2})^{2}}$, where $\Delta
T=T_{L}-T_{R}$ and $\bar{T}=(T_{R}+T_{L})/2$. The Wiedemann-Franz law
holds and is given by $K/(G\bar{T})=\pi^{2}/3\equiv L_{0}$, the Lorentz
number.

In general, if bias voltages and temperature gradients are applied,
the electric and heat currents are given by
\begin{eqnarray}
I & = & L_{11}V+L_{12}\Delta T+I_{0},\nonumber \\
I_{Q} & = & L_{21}V+L_{22}\Delta T.
\label{eq:TransportCoefficients}
\end{eqnarray}

For the SJ case, $L_{11}=G$, $L_{22}=K$ and $L_{12}=L_{21}=0$; so no
thermoelectric effects are observed.  For the AJ case, expanding the
result to linear order in $V$ and $\Delta T$, we see that still
$L_{12}=L_{21}=0$. (Notice that thermoelectric effects are observed in
this case, but not at the linear order.) In terms of the transport
coefficients, the Lorentz number is defined as $L_{22}/(L_{11}
\bar{T})$. Our calculations show deviations from the Wiedemann-Franz law.
This behavior is due to the same effects as in the case of
the conductance and the presence of the nonzero $I_{0}$ for the AJ case; 
$L_{0} \simeq \pi^{2}/3$ for small tunneling
strengths, but it increases as the strength increases.

We have presented a general scheme to deal with non-Hermitian
Hamiltonians, and have shown that the generalized Schwinger-Keldysh formalism is a natural
choice for such applications.  As an example, we have studied a simple
model of an asymmetric tunneling junction where irreversibility is
explicitly encoded into the Hamiltonian, hence making it non-Hermitian.  We have
calculated the steady state transport properties of the junction and seen that the
behavior of the observables goes along with our physical
expectations for the model. The approach is not limited to systems with steady
states and can be used to address general non-equilibrium
situations. We intend to apply it, for instance, to study the formation
and evaporation dynamics of interacting BECs, where a steady state is
not achieved, and explore regimes (\emph{e.g.}, the unitary limit) that are not accessible with present
methods.

\begin{acknowledgments}
We would like to acknowledge discussions with J. Kono and
A. Srivastava which inspired this letter. This work was partly
supported by the W. M. Keck Foundation and DARPA/ARO (W911NF-07-1-0464).
\end{acknowledgments}

\bibliographystyle{apsrev}
\bibliography{paper}

\end{document}